\documentclass[twocolumn,showpacs,preprintnumbers,amssymb]{revtex4}

\usepackage{graphicx}
\usepackage{dcolumn}
\usepackage{bm}
\usepackage{dsfont}
\usepackage{amsmath}
\usepackage{pifont}
\usepackage{psfrag}
\usepackage{pstricks}
\usepackage[latin2]{inputenc}

\newcommand\bra[1]{\langle{#1}|}
\newcommand\ket[1]{|{#1}\rangle}

\newcommand\ibrkt[2]{|{#1}\rangle\langle{#2}|}

\newcommand{\be}{\begin{equation}}
\newcommand{\ee}{\end{equation}}

\newcommand{\ehoch}[1]{e^{#1}}
\newcommand{\skipc}[2]{}

\newcommand{\fig}[1]{Fig.~\ref{#1}}
\newcommand{\eq}[1]{Eq.~(\ref{#1})}

\begin{document}

\preprint{APS/123-QED}

\title{Chirping a two-photon transition in a multi-state ladder}

\author{Wolfgang Merkel}
\author{Holger Mack}
\author{Wolfgang P. Schleich} %
\email{wolfgang.schleich@uni-ulm.de}
\affiliation{%
Institut f\"ur Quantenphysik, Universit\"at Ulm,\\ Albert-Einstein-Allee 11, D-89081 Ulm, Germany 
}
\author{Eric Lutz}
\affiliation{Department of Physics, Universit\"at Augsburg,
D-86135 Augsburg, Germany
}
\author{Gerhard G. Paulus}
\affiliation{Department of Physics, Texas A\&M University, 
College Station, TX 77843, United States
}
\author{Bertrand Girard}
\affiliation{Laboratoire de Collisions, Agr\'egats et R\'eactivit\'e, (UMR 5589, CNRS-Universit\'e Paul Sabatier Toulouse 3), IRSAMC, Toulouse, France
}

\date{\today}

\begin{abstract}
We consider a two-photon transition in a specific ladder system driven by a chirped laser pulse. In the weak field limit, we find that the excited state probability amplitude arises due to interference of multiple quantum paths which are weighted by quadratic phase factors. The excited state population has the form of a Gauss sum which plays a prominent role in number theory.
\end{abstract}


\maketitle

\section{Introduction}
Fresnel diffraction at a straight edge leads to an intensity pattern determined by the Cornu spiral\cite{born:wolf,chirp:footnote}.
The latter results from an integral with a quadratic phase. The Talbot effect\cite{talbot:1836} is a generalization of this scattering situation.
The discreteness of the grating translates the integral of quadratic phase factors into a sum. Recently, the combination of chirped pulses together with an appropriate atomic level scheme created an atomic analog of the Fresnel diffraction of the straight edge\cite{noordam:diffraction:1992,girard:2001}. In the present paper we propose the atomic analog of the Talbot effect by chirping a two-photon transition in a multi-state ladder.

Chirped pulses are characterized by a non-linear phase dependence and have a vast variety of applications\cite{noordam:diffraction:1992,noordam:1994,noordam:1999,Corkum1999,gerber:1996,girard:optcom:2006}. 
Examples include, the realization of a time-domain Fresnel lens with coherent control \cite{girard:2002} and quantum state measurement using coherent transients\cite{girard:2006}.
Of particular relevance in the context of the present paper are interference phenomena in quantum ladder systems induced by two-photon excitation with chirped laser pulses\cite{girard:2003,girard:2004}.

\begin{figure}
\includegraphics[width=\columnwidth]{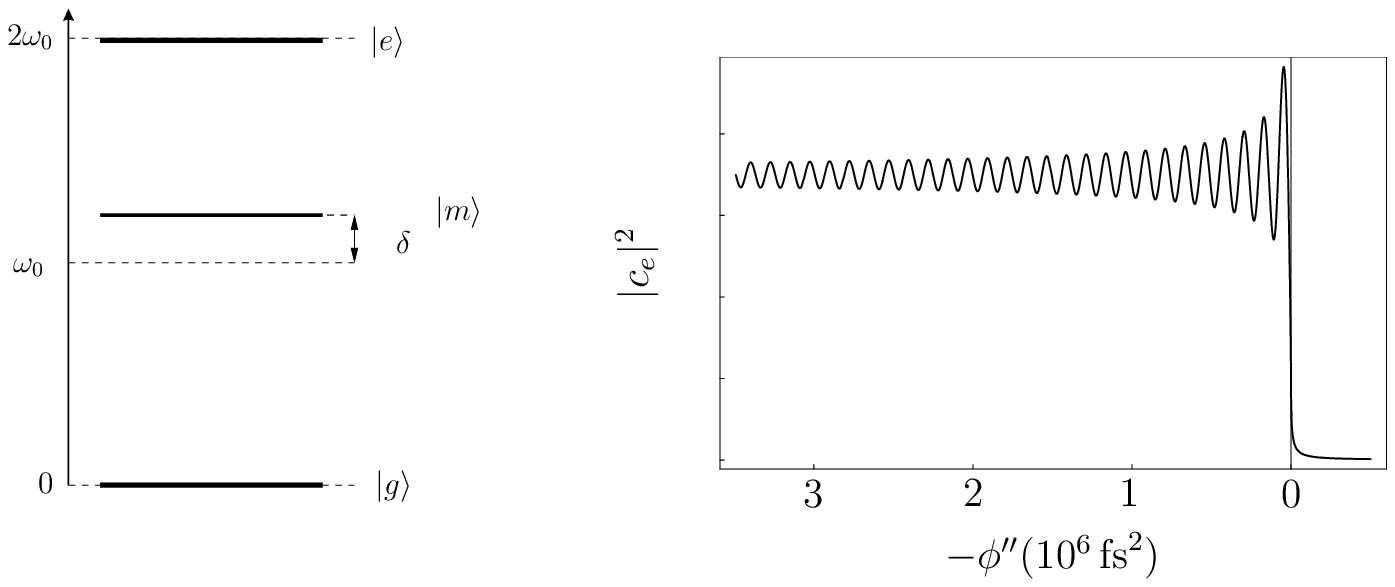}
\\[5mm]
\includegraphics[width=\columnwidth]{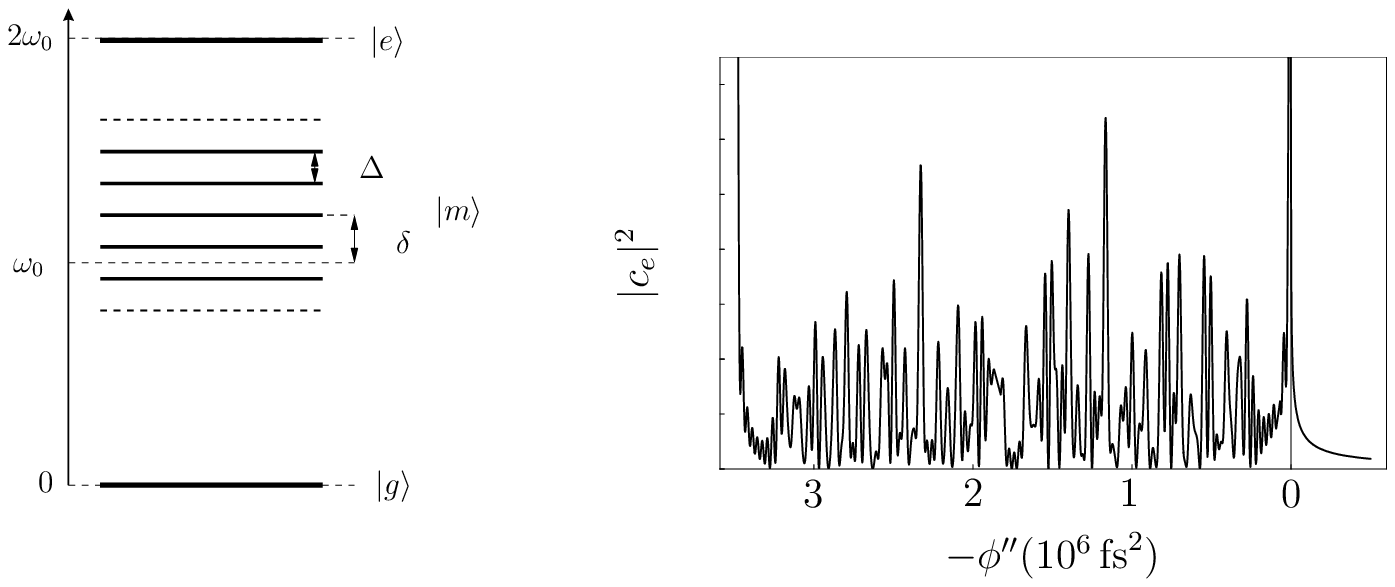}
\caption{Chirping a two-photon excitation through a single intermediate state (top) and through an equidistant manifold (bottom).
The left and right columns show the level system and the resulting excitation probabilities as a function of the chirp parameter $\phi''$, respectively. In both systems the  ground state $\ket{g}$ is connected by a two-photon transition to the excited state $\ket{e}$. 
In the case of a single intermediate state $\ket{m}$ with the offset $\delta>0$ with respect to the central frequency $\omega_0$  discussed  in the top row the population $|c_e|^2$ of the excited state displays oscillations for negative chirps, whereas for positive values of the chirp it is decreasing rapidly. The oscillations result from an interference of the direct and the indirect excitation. The direct path is a direct two-photon transition through the virtual state of frequency $\omega_0$. The indirect path consists of first an excitation to the level $\ket{m}$ and then a transition to $\ket{e}$. 
Here we have used the parameters 
$
\delta=0.0225\, \text{fs}^{-1},\,
\Delta\omega= 0.1525\,  \text{fs}^{-1},\,
\Omega_{em}\Omega_{mg}=1
$
as suggested by the experiment\cite{girard:2003}.
In the bottom row we discuss the case of a manifold of $M'+M+1$ equidistant intermediate states $\ket{m}$ with $-M'\le m \le M$ which are shifted by the offset $\delta_m=\delta+m\,\Delta$ with respect to the central frequency $\omega_0$. The offset of the central state is $\delta$ whereas the states in the harmonic manifold are separated by $\Delta$.
The population $|c_e|^2$ of the excited state is due to interference of competing quantum paths via the intermediate states and displays strong modulations as a function of the chirp. 
Parameters are
$M'=7,\,
M=7,\, 
\delta=0.0225\, \text{fs}^{-1},\,
\Delta=0.003\,\text{fs}^{-1},\,
\Delta\omega= 0.1525\,  \text{fs}^{-1},\,
\Omega_{em}\Omega_{mg}=1$
}
\label{fig1}
\end{figure}

The central idea is summarized in \fig{fig1}. We consider a two-photon transition from the ground state $\ket{g}$ to the excited state $\ket{e}$.
In the case of a single intermediate state\cite{chirp:footnote:2,greenland} discussed in the top row of this figure the total excitation probability  $W_e\equiv|c_e|^2$ is due to the interference of two paths: ({\it i}) in the direct path two photons of energy $\hbar \omega_0$ are absorbed instantaneously and ({\it ii}) in the  sequential excitation path the total energy of $2\hbar\omega_0$ is absorbed in two steps: in the first step the system is excited to the intermediate state $\ket{m}$, whereas the second, delayed step provides the residual energy  to reach the excited state $\ket{e}$. This interference effect gives rise to an excitation probability determined by a Fresnel integral. It is experimentally accessible by measuring the population in the excited state through the detection of the fluorescence signal. Population transfer with chirped laser pulses has been experimentally demonstrated in the three-state ladder of Rubidium \cite{noordam:1994}. 

In the case of a multi-state ladder with a manifold of intermediate states shown in the second row of \fig{fig1} we find a generalization of this two-path interference to a multi-path interference.
Here the interference is mainly due to the indirect paths through the individual intermediate states. The direct path does not play a dominant role anymore. The problem of scattering a chirped pulse from $D$ intermediate levels is very much in the spirit of diffraction from a $D$-slit grating. For an equidistant manifold  the phase of the excitation path through the $m$-th level contributing to the total excitation probability depends quadratically on $m$ in complete analogy to the Talbot effect. We emphasize that the present proposal is different from the temporal Talbot effect suggested in Ref.~\cite{mitschke:1998}. 

Our paper is organized as follows: 
In Sec.~\ref{model}, we investigate the population transfer in the $D$--state ladder system shown in \fig{fig1} driven by a weak chirped laser pulse. We use second order perturbation theory to derive  an analytic expression for the probability amplitude to be in the excited state. This result is of the form of a Gauss sum\cite{davenport:1980,schleich:2005:primes} which plays a prominent role in number theory. 
We dedicate the following sections to a discussion of the physical origin of this Gauss sum. 

We start in Sec.~\ref{single:intermediate:state} by first considering the transition probability amplitude $c_e$ in the limit of a single intermediate state. In this case $c_e$ follows from the complementary error function with a complex-valued argument which depends on the chirp parameter. With the help of appropriate asymptotic expansions of the complementary error function we bring to light the oscillations in the transition probability $W_e=|c_e|^2$ shown in \fig{fig1} for negative values of the chirp. These oscillations arise due to the interference of two excitation paths as shown in Sec.~\ref{interference}. Moreover, a primitive version of the method of stationary phase allows us to  identify the times of the transitions. We show that the accumulated phases depend quadratically on the offset $\delta_m$. Since our analysis works in frequency-time phase space our approach is different from Refs.~\cite{silberberg:2001,girard:2003,girard:2004}
which relies on the frequency domain. In the case of many intermediate levels the interference of quadratic phase factors gives rise to a Gauss sum which we express in its canonical form in Sec.~\ref{manifold}. Here all dimensionless quantities are related to experimental parameters. 
In Sec.~\ref{conclusion} we conclude by addressing possible experimental realizations of multi-state systems.

\section{Excitation probability}
\label{model}
We consider the multi-state ladder system shown in the bottom of \fig{fig1}
driven by a weak chirped laser pulse and
calculate the probability amplitude of the excited state in second order perturbation theory.  We show that this probability amplitude is the interference of many amplitudes which depend quadratically on the offset of the intermediate state. We present a simple phase space argument for the origin of such quadratic phase factors.

\subsection{Model system}
The  ladder system consists of a ground state $\ket{g}$, an excited state $\ket{e}$ separated by an energy $2\hbar\omega_0$ and intermediate states $\ket{m}$ with quantum numbers $M'\le m \le M$, as depicted in \fig{fig1}. These intermediate states are displaced by the offset
\begin{equation}
\label{energy}
\delta_m=\delta+m\,\Delta.
\end{equation}
with respect to the central frequency $\omega_0$.
The harmonic manifold is characterized by the offset $\delta$ of a specific central state $\ket{0}$ and separation $\Delta$ of adjacent states.

In the interaction picture the  Hamiltonian  describing the interaction between the electric field of the chirped pulse
\be
E(t)={\cal E}_0\left[ \ehoch{-i\omega_0 t}\,f(t)+\text{c.c.}\right]
\label{eq:el:field}
\ee
of amplitude ${\cal E}_0$, carrier frequency $\omega_0$ and  pulse shape function $f(t)$ and our ladder system reads 
\begin{equation}
\begin{split}
V(t)=-\hbar\sum\limits_m \Big[\Big.&
\Omega_{mg}\,  \ehoch{i\,\delta_m t } f(t)\,\ibrkt{m}{g}
+\\
&\Omega_{em}\, \ehoch{-i\,\delta_m t} f(t)\,\ibrkt{e}{m}+
\text{c.c.}
\Big.\Big]
\end{split}
\label{Hint}
\end{equation}
where we have introduced the  Rabi frequencies $\Omega_{ij}\equiv
{\cal E}_0\wp_{ij}/\hbar$ with the dipole matrix element $\wp_{jm}$  of the transition $\ket{j} \to \ket{m}$ and $\ket{j} =\ket{g,e}$.

\subsection{Time evolution}
Throughout the paper we consider a situation in which the laser pulse is short (femtosecond--time scale) compared to the characteristic time scale of radiative decay processes. As a consequence we can neglect spontaneous emission and the time-evolution of the state vector 
\be
\label{eq4}
\ket{\psi(t)}=\sum\limits_k c_k(t)\,\ket{k}
\ee
is given by the Schr\"odinger equation
\be
i\hbar\frac{d}{d t}\ket{\psi(t)}=V(t)\ket{\psi(t)}.
\label{schroedi}
\ee
Since the laser field is assumed to be weak,
the probability amplitude $c_e$ of the excited state can be simply calculated in second order perturbation theory. We find
\be
c_e(t)=-\frac{1}{2\hbar^2}\int\limits_{-\infty}^t d t'\!
 \int\limits_{-\infty}^{t'}d t''\;\bra{e}V(t')\,V(t'')\ket{g}
\label{eUg}
\ee
The first--order term vanishes since
we assume that at $t_0=-\infty$, that is before the interaction, the ladder system is initially prepared in the ground state $\ket{g}$ and a two-photon transition is necessary in order to reach the excited level.

When we substitute the
explicit expressions for the interaction Hamiltonian, \eq{Hint},  and concentrate on times after the pulse has passed, that is $t\to\infty$, we obtain
\begin{equation}
\label{eUg3}
c_e=-\frac{1}{2} \sum\limits_m \Omega_{em}\Omega_{mg}\, {\cal I}_m
\end{equation} 
for the probability amplitude $c_e=c_e(t\to \infty)$.

The domain of the integral
\be
{\cal I}_m\equiv
\int\limits_{-\infty}^\infty dt'\int\limits_{-\infty}^{t'} dt''\,e^{-i\,\delta_m (t'-t'')}f(t')f(t'')
\label{I:m}
\ee
extends over the area below the  diagonal crossing  the  $(t',t'')$-plane from the lower left to the upper right. When we introduce the new integration variables  $\bar{t} \equiv t'+t''$ and $t \equiv t'-t''$  and use the formula 
\be
\int\limits_{-\infty}^\infty dt'\,\int\limits_{-\infty}^{t'} dt''\dots =\frac{1}{2}
\int\limits_{-\infty}^\infty d\bar{t}\,\int\limits_{0}^{\infty} dt \dots
\label{timechange}
\ee
we eventually arrive at
\begin{equation}
\begin{split}
c_e=&-\frac{1}{4} \sum\limits_m \Omega_{em}\Omega_{mg}\times\\
&\int\limits_{-\infty}^\infty d\bar{t}\,
\int\limits_{0}^{\infty} dt \,
\ehoch{-i\,\delta_m t} \,
f\left(\frac{\bar{t}+t}{2}\right)\,
f\left(\frac{\bar{t}-t}{2}\right)
.
\end{split}
\label{exprob}
\end{equation} 
This expression is valid for an arbitrary pulse shape $f$. In the next section we restrict ourselves to a linearly chirped pulse with a Gaussian envelope.

\subsection{Description of the chirped pulse}

We now specify the pulse shape 
\begin{equation}
\label{pulse}
f(t)=f_0 \,
\exp\left[ -\frac{1}{2}(\Delta\omega f_0)^2 t^2\right]
\end{equation}
with the complex-valued amplitude
\be
f_0=\sqrt{\frac{1+ia}{1+a^2}},
\label{f0}
\ee
and the dimensionless parameter $a$ denotes the second order dispersion
\be
a=\Delta\omega^2 \phi''.
\label{scaledchirp}
\ee
Here $\Delta\omega$ is the bandwidth of the pulse and
$\phi''={d^2} \phi(\omega)/{d\omega^2}$
is a measure for the quadratic frequency dependence of the phase of the laser pulse. 

It is useful to decompose the argument 
\be
\frac{1}{2}(\Delta\omega f_0)^2\equiv \alpha_r+i \,\alpha_i
\label{f0vonalpha}
\ee
of the Gaussian pulse shape given by \eq{pulse} into 
the real and imaginary parts
\be
\alpha_r \equiv \frac{\Delta\omega^2}{2}\frac{1}{1+a^2}
\quad \text{and} \quad
\alpha_i \equiv \frac{\Delta\omega^2}{2}\frac{a}{1+a^2}
\label{alphas}
\ee
resulting in
\be
f(t)= f_0\, \exp\left(-\alpha_r t^2\right)\,\exp\left(-i\,\alpha_i t^2\right).
\ee
This representation brings out most clearly that the electric field \eq{eq:el:field} of the chirped laser pulse  features a linear variation of the instantaneous frequency
\begin{equation}
\omega(t)
=\omega_0+2\alpha_i\,t.
\label{instantaneous}
\end{equation}

\subsection{Excitation probability amplitude}
When we now substitute the pulse shape, \eq{pulse}, into the expression \eq{exprob} for the excitation probability amplitude we obtain 
\begin{equation}
\label{eq21}
c_e=-\frac{1}{4}\sum\limits_m \Omega_{em}\Omega_{mg}\, I_1\, I_2
\end{equation}
with 
\be
I_1\equiv 
f_0\int\limits_{-\infty}^{\infty}d \bar{t}\;
e^{-\frac{1}{4}(\Delta\omega f_0)^2\, \bar{t}^2}
=\frac{2\sqrt{\pi}}{\Delta\omega}
\ee
and
\be
I_2\equiv f_0\int\limits_{0}^{\infty}d t\;
e^{-i\delta_m t}
e^{-\frac{1}{4}(\Delta\omega f_0)^2\, t^2}.
\label{I2:t:integration}
\ee 
The substitution 
\be
z\equiv \frac{\Delta\omega}{2} f_0\, t+ i\, \frac{\delta_m}{\Delta\omega f_0}
\label{subst}
\ee
casts the integral $I_2$ into the form 
\be
I_2=\frac{\sqrt{\pi}}{\Delta\omega} 
\exp\left[-\left(\frac{\eta_m}{f_0}\right)^2\right]
\text{erfc}\left(i \frac{\eta_m}{f_0}\right)
\ee
where we have introduced the dimensionless offset
\be
\eta_m\equiv \frac{\delta_m}{\Delta\omega}.
\ee
and have recalled\cite{abramowitz} the definition 
\be
\text{erfc}(\zeta) \equiv \frac{2}{\sqrt{\pi}}\int\limits_{{\cal C}(\zeta)} dz\; \ehoch{-z^2}
\label{erfc:definition}
\ee
of the complementary error function. 

The integration in \eq{I2:t:integration} over $t$ from $0$ to $\infty$ translates into a path ${\cal C}(\zeta_m)$ in the complex plane which according to \eq{subst} starts at the point $\zeta_m \equiv i\,\eta_m/f_0$ and follows a straight line to infinity. This path encloses an angle $\text{arg}(f_0)$ with respect to the positive real axis.

When we introduce the abbreviation 
\be
d_m \equiv \pi \frac{\Omega_{em}\Omega_{mg}}{\Delta\omega^2}
\label{dm}
\ee
the definition \eq{f0} of $f_0$ yields the expression  
\be
c_e=
\sum\limits_m \,w_m\,
\exp\left(i\,\eta_m^2 a\right)
\label{ce}
\ee
for the transition probability amplitude \eq{eq21}. Here we have introduced the weight factors 
\be
w_m\equiv-\frac{d_m}{2}\, 
\text{erfc}\left(\zeta_m(a)\right)
\exp{\left(-\eta_m^2\right)}
\label{eq:weight:1}
\ee
with 
\be
\zeta_m(a)=\eta_m\, i\, \sqrt{1-i\,a}.
\ee
Equation~(\ref{ce}) is the central equation of the present paper. It demonstrates that the probability amplitude $c_e$ to be in the excited state is a sum of weighted phase factors in which the summation index $m$ enters through the dimensionless offset $\eta_m$. Since  according to \eq{energy} the offset $\delta_m$ is linear in $m$ and the phase is quadratic in $\eta_m$ the phase factors are quadratic in $m$. Such sums are called Gauss sums\cite{schleich:2005:primes,davenport:1980}.

\section{Single intermediate state}
\label{single:intermediate:state}
In order to gain some insight into the dependence of the transition probability amplitude $c_e$ on the dimensionless chirp parameter $a$ we first concentrate on a single intermediate state $\ket{m}$. In this case the sum in  \eq{ce} reduces to a single term and the transition probability 
\be
W_e=|c_e|^2=\frac{1}{4}\,d_m^2\,\ehoch{-2\eta_m^2}\,
\left|\text{erfc}\left(\zeta_m(a)\right)\right|^2
\label{zeta}
\ee
is solely determined by the weight factor $w_m$. As a consequence the quadratic phase $\eta_m^2a$ in \eq{ce} has no chance to contribute. Nevertheless, under appropriate conditions\cite{noordam:1994,noordam:1992,girard:2003,girard:2004} we find interference due to the rather intricate properties of the complementary error function. In order to bring this subtle phenomenon to light we now analyze the asymptotic behavior of $\text{erfc}$. 

\subsection{Path of integration induced by chirp}

According to \eq{zeta} the transition probability $W_e$ follows from the absolute value squared of the complementary error function evaluated at the argument $\zeta_m(a)$.
The function $\text{erfc}(\zeta_m)$ is an integral  of the analytic function 
\be
\ehoch{-z^2}=\ehoch{-(z_r^2-z_i^2)}\,\ehoch{-i\,2\,z_r\,z_i}
\ee
along the path ${\cal C}(\zeta_m)$.
For $|z_i|< |z_r|$  the integrand $\exp(-z^2)$ decreases rapidly for increasing $|z_r|$. The decaying domains for positive and negative real parts are connected via a saddle at the origin. Indeed, in the sector $|z_r|<|z_i|$ we deal with a rapidly growing integrand as $|z_i|$ increases.

As a function of the chirp parameter $a$ the starting point $\zeta_m(a)$ of the integration traverses a path in complex plane. When we  square the representation 
\be
\zeta_m \equiv \eta_m(x+i\,y)
\ee
and take the real and imaginary parts of  
\be
x^2-y^2+2i\,xy=i\, a-1,
\ee
we find the trajectory
\be
y=\sqrt{1+x^2}
\label{trajectory1}
\ee
and the translation 
\be
a=2xy=2x\,\sqrt{1+x^2}
\label{eq2}
\ee
between $x$ and $a$. 

For large positive or large negative values of $x$ the trajectory, \eq{trajectory1}, of $\zeta_m$  approaches the diagonals of the first or second quadrant, respectively. The straight path ${\cal C}$ of integration defined in \eq{subst} has a steepness determined by the coefficient \be
\frac{1}{2}\Delta\omega f_0=\frac{\Delta\omega}{2\sqrt{1-i\,a}}= \frac{\Delta\omega}{2} (1+a^2)^{-1/4} \ehoch{\frac{i}{2}\arctan(a)}
\ee
in front of $t$.
Hence, for large positive values of $a$ the straight path encloses an angle with respect to the positive real axis which is slightly smaller than $\pi/4$. For $a=0$ the path is parallel to the positive real axis. For large negative values of $a$ the angle of inclination of ${\cal C}$ is slightly larger than $-\pi/4$.

\begin{figure}[]
\begin{center}
\includegraphics[width=0.75\columnwidth]{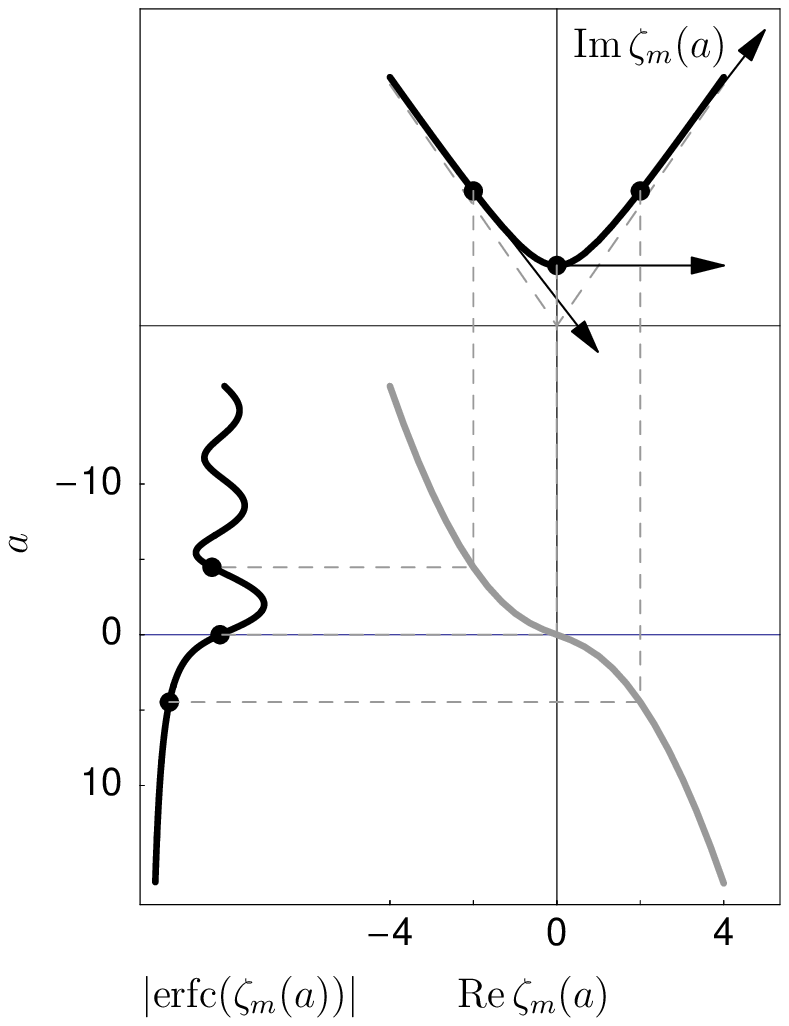}
\end{center}
\caption{
Complementary error function in its dependence on the chirp parameter $a$  (left bottom)  as obtained by integrating along a straight path in complex space (top). The starting point $\zeta_m(a)$ of this path traverses as a function of $a$ a trajectory in complex space which for large positive and large negative values of $a$ approaches the diagonals of the first and the second quadrants, respectively. The path is always a straight line which is tangent to this trajectory and tends  towards infinity as shown on the top. The gray curve below translates between $a$ and $\text{Re}\,\zeta_m$.  
}
\label{chirp:trajectory}
\end{figure}

In \fig{chirp:trajectory} we show on the top by a solid line the trajectory $y(x)$ given by \eq{trajectory1} together with the direction of the path ${\cal C}$ indicated by arrows. Below we depict the translation function $a(x)$ connecting the chirp parameter $a$ with the real part of $\zeta_m$. The picture on the bottom left shows the absolute value of the complementary error function $\text{erfc}(\zeta_m(a))$ as a function of $a$.

\subsection{Asymptotic expressions}
\label{asymp}
The top of \fig{fig1} and the bottom left of \fig{chirp:trajectory} demonstrate that the value $a=0$ of the chirp parameter separates two distinct domains of the transition probability: ({\it i}) For positive values of $a$  $W_e$ decays as $a$ increases, and ({\it ii}) for negative values of $a$ the probability oscillates. These qualitatively different behaviors of $W_e$ originate from the different slopes of the paths of integration as we now show. 

\subsubsection{Decaying regime}
We start our discussion by analyzing the behavior of $\text{erfc}(\zeta_m(a))$  for large positive values of $a$. For this purpose we integrate the integral in the definition \eq{erfc:definition} by parts which yields
\be
\text{erfc}(\zeta_m)= \frac{1}{\sqrt{\pi}}\left[\left.-\frac{\ehoch{-z^2}}{z}\right|_{{\cal C}_+}+ \int\limits_{{\cal C}_+}dz\; \frac{\ehoch{-z^2}}{z^2}\right].
\label{A3}
\ee
Since the path ${\cal C}_+$ shown in \fig{erfc:int:path} reaches the domain where $\ehoch{-z^2}$ decays the boundary term originating from the end point of the path ${\cal C}_+$, that is from infinity, vanishes. As a consequence only the starting point of ${\cal C}_+$ contributes leading to 
\be
\text{erfc}(\zeta_m)= \frac{1}{\sqrt{\pi}}\frac{\ehoch{-\zeta_m^2}}{\zeta_m}+
\int\limits_{{\cal C}_+} dz\;\frac{\ehoch{-z^2}}{z^2}.
\ee
We can now repeat this procedure of partial integration to obtain a power series of $\text{erfc}$ in $1/\zeta_m$. In lowest order we arrive at the asymptotic expression 
\be
\text{erfc}(\zeta_m)\stackrel{\sim}{=} \frac{1}{\sqrt{\pi}}\frac{\ehoch{-\zeta_m^2}}{\zeta_m}
\label{erg1}
\ee
valid for $1 \ll a$.

\subsubsection{Oscillatory regime}

We  now turn to the limit of large negative values of $a$. It is tempting to apply again the technique of integration by parts. However, this idea is bound to fail since the partial integration creates inverse powers of $z$ in the integrand. As shown in \fig{chirp:trajectory} for large negative values of $a$ the path of integration gets very close to the origin of complex space. As a result the series cannot converge in this case. This heuristic argument clearly indicates that the behavior for negative values of $a$ is substantially different from the one arising for $0<a$.

\begin{figure}
\begin{center}
\includegraphics[width=\columnwidth]{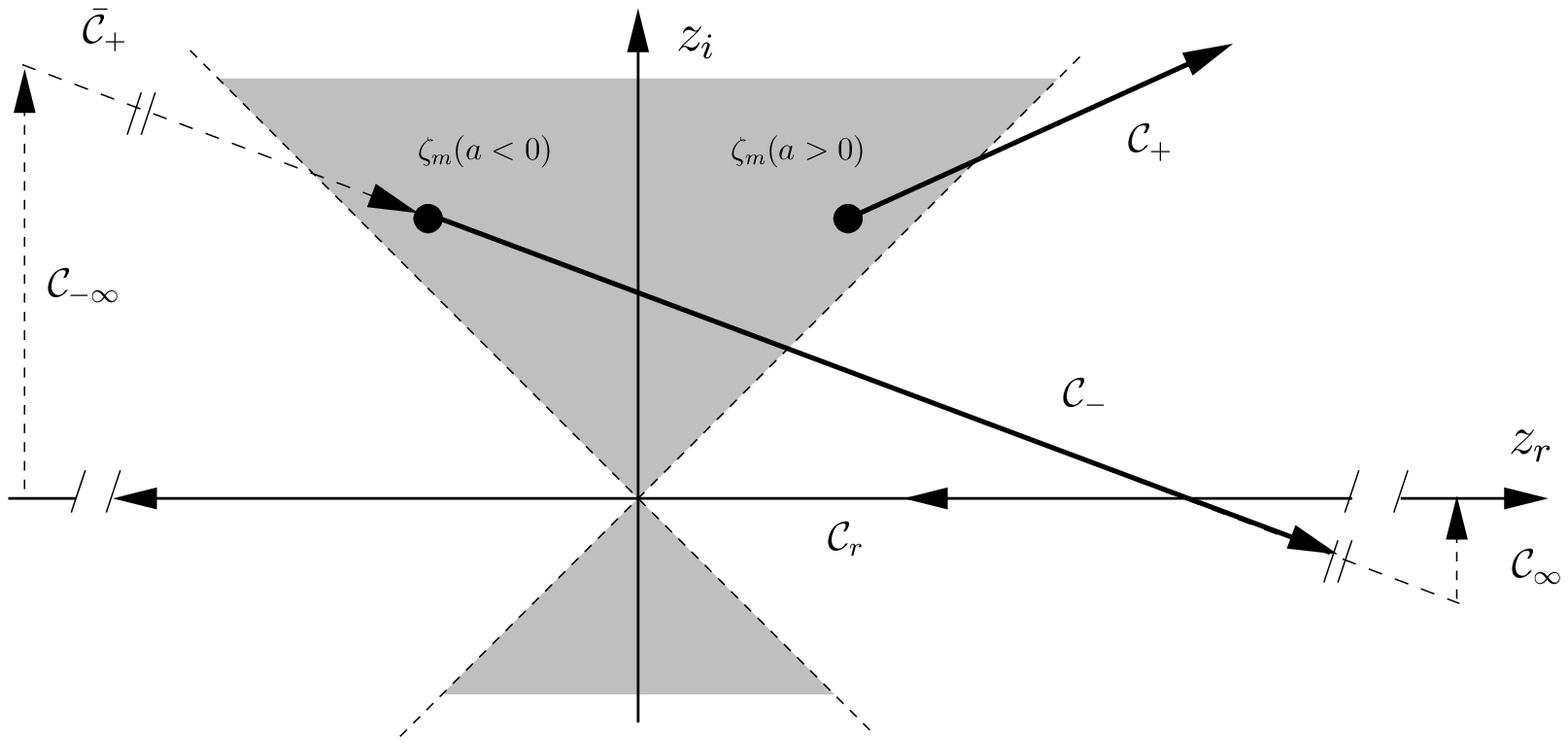}
\end{center}
\caption[Integration paths in complex plane]
{
Paths ${\cal C}_+$ and ${\cal C}_-$ of integration in the complex plane corresponding to a positive and a negative value of the chirp parameter $a$ .
We depict the domains in the complex plane where the integrand  $\ehoch{-z^2}$ of the complementary error function increases by shaded areas. 
For $0<a$ the starting point $\zeta_m$ of the integration is located in the first quadrant and the straight path ${\cal C}_+$ is inclined with a steepness slightly smaller than the first diagonal. Thus ${\cal C}_+$ proceeds almost completely in the domain of the complex plane where the integrand $\exp(-z^2)$ of $\text{erfc}$ decays.
For $a<0$ the path ${\cal C}_{-}$ of integration starts in the second quadrant and proceeds slightly above the second diagonal into the decaying region of $e^{-z^2}$.
In order to evaluate the resulting integral we close the path as indicated by the dashed line consisting of the paths ${\cal C}_{-}$,  ${\cal C}_{\infty}$, ${\cal C}_{r}$, ${\cal C}_{-\infty}$  and $\bar{{\cal C}}_{+}$.}
\label{erfc:int:path}
\end{figure}

In order to evaluate $\text{erfc}(\zeta_m(a))$ for large negative  values of $a$ we first recall that $\ehoch{-z^2}$ is an analytic function. Therefore, the Cauchy theorem applied to the closed path ${\cal C}_c$ depicted in \fig{erfc:int:path} yields
the identity
\be
\oint\limits_{{\cal C}_c} dz\; \ehoch{-z^2}=0.
\ee 
Since in infinity the integrand vanishes the contributions from the paths ${\cal C}_\infty$ and ${\cal C}_{-\infty}$ are zero and we find immediately
\be
\int\limits_{{\cal C}_{-}} dz\; \ehoch{-z^2}=
\sqrt{\pi}-
\int\limits_{\bar{{\cal C}}_+} dz\; \ehoch{-z^2}.
\ee
The path $\bar{{\cal C}}_+$ is the mirror image of the path ${\cal C}_+$ and avoids the origin. Therefore, we can use the technique of the preceding section to approximate the remaining integral. When we traverse $\bar{{\cal C}}_+$ in the opposite direction we arrive at
\be
\text{erfc}(\zeta_m)\stackrel{\sim}{=}
2+\frac{1}{\sqrt{\pi}}\frac{\ehoch{-\zeta_m^2}}{\zeta_m}.
\label{erg2}
\ee
In this approximation the complementary error function is a sum of two complex-valued contributions: a constant term and a phase factor. As a result the interference between these two terms  manifests itself in oscillations of the probability $W_e$  as a function of $a$.  

\section{Interference of excitation paths}
\label{interference}
The preceding section has shown that the two paths $\bar{{\cal C}}_+$ and ${\cal C}_r$ of integration in complex space correspond to two interfering contributions in the total probability amplitude $c_e$. We now show that they represent two different paths of excitation in the atom. Indeed, the term $\ehoch{-\zeta_m^2}/\zeta_m$ results from a direct two-photon transition whereas the term 2 results from a sequential excitation.

\subsection{Sequential excitation}
So far our analysis was based on the integral $I_2$. However, in order to identify the sequential character of the excitation process it is more convenient to use the representation \eq{eUg3} for the transition probability amplitude $c_e$ with  the integral ${\cal I}_m$. The latter is the product of two integrals which  reflects the sequential character of the excitation process: in the first step, the atom is promoted to the intermediate state $\ket{m}$, while  the second step provides the residual energy to reach the excited state. 

\subsubsection{Times of excitation}

In order to bring out this feature most clearly and to identify the times when the transitions occur we now substitute the expression \eq{pulse} for the shape $f(t)$ of the pulse into the definition \eq{I:m} of ${\cal I}_m$ which leads us to the integral 
\begin{eqnarray}
{\cal I}_m
&=&f_0^2 \ehoch{i\, \frac{\delta_m^2}{4\alpha_i}}
\int\limits_{-\infty}^\infty dt'\;\ehoch{-\alpha_r t'^2} \ehoch{-i\,\alpha_i\left(t'+\frac{\delta_m}{2\alpha_i}\right)^2}\nonumber
\\&&\times
\ehoch{i\, \frac{\delta_m^2}{4\alpha_i}}
\int\limits_{-\infty}^{t'} dt''\;\ehoch{-\alpha_r t''^2} \ehoch{-i\,\alpha_i\left(t''-\frac{\delta_m}{2\alpha_i}\right)^2}.
\label{quad:erg}
\label{int:timeorder}
\end{eqnarray}
So far the calculation is exact. We now recognize that for large values of the chirp parameter $a$ the definition \eq{alphas} of $\alpha_r$ and $\alpha_i$ provides us with the inequality
\be
\left|\frac{\alpha_r}{\alpha_i}\right|=\frac{1}{|a|}\ll 1.
\ee
Consequently, the real-valued Gaussians in the integral \eq{quad:erg} are slowly varying compared to the oscillatory terms resulting from the quadratic phases. The main contributions to the integral emerge from the oscillatory terms and in particular from the neighborhood of the times 
\be
t_s''\equiv \frac{\delta_m}{2\alpha_i}=\frac{\eta_m}{\Delta\omega}a\left(1+\frac{1}{a^2}\right)\stackrel{\sim}{=} \frac{\eta_m}{\Delta\omega} a
\label{eq39}
\ee
and
\be
t_s'=-\frac{\delta_m}{2\alpha_i}=-t_s''
\ee
when the phase factors are slowly varying.

Therefore, the transition from the level $\ket{g}$ to the level $\ket{m}$ appears at the time $t''=t_s''$ followed by the transition from $\ket{m}$ to $\ket{e}$ at the time $t'=t_s'$. In order to ensure this time ordering enforced by the limits $t''<t'$ of the integral ${\cal I}_m$ defined in \eq{quad:erg} we need to have 
\be
t_s''<t_s'=-t_s''.
\label{inequality}
\ee

This inequality is only satisfied provided $t_s''$ is negative. From \eq{eq39} we recognize that it is the product $\eta_m a$  which determines the sign of $t_s''$. Consequently we only find points of slowly varying phase for $\eta_m a<0$. In the case $0<\eta_m a$ no such points exist and the integral ${\cal I}_m$ is decreasing rapidly.

\subsubsection{Origin of quadratic phases}

Moreover, this analysis of the integral ${\cal I}_m$ brings out most clearly that each transition is associated with a phase which is quadratic in the dimensionless offset $\eta_m$. The phases of both transitions are identical. The total acquired phase in the two-photon transition is twice of that of the individual one-photon transition.

We can now evaluate the integrals in an approximate way by replacing the variables $t'$ and $t''$ in the real-valued Gaussians by $t_s'$ and $t_s''$ and factoring them out of the integral which yields
\begin{eqnarray}
{\cal I}_m 
&\stackrel{\sim}{=}& 
\Theta(-\eta_m a) f_0^2 
\left(\int_{-\infty}^\infty du\; \ehoch{-i\,\alpha_i u^2}\right)^2 \nonumber\\
&&\times
\exp\left[-\frac{\alpha_r}{\alpha_i} \frac{\delta_m^2}{2\alpha_i}\right]
\exp\left(i\, \frac{\delta_m^2}{2\alpha_i}\right).
\end{eqnarray}
Here we have also extended the integration over $t''$ to $+\infty$.

With the help of the integral relation\cite{abramowitz}
\be
\int\limits_{-\infty}^\infty du\; \ehoch{-i\, \gamma u^2}=
\sqrt{\frac{\pi}{i\,\gamma}}
\ee 
we find
\be
{\cal I}_m \approx \Theta(-\eta_m a)
f_0^2
\frac{\pi}{i\,\alpha_i}
\exp\left[-\frac{\alpha_r}{\alpha_i}\frac{\delta_m^2}{2\alpha_i}\right]
\exp\left[i\,\frac{\delta_m^2}{2\alpha_i}\right].
\ee
When we now take the limit of $1\ll |a|$ in the definitions \eq{f0} and \eq{alphas} of $f_0$ and $\alpha_i$ which yields $f_0^2\stackrel{\sim}{=} i/a$ and $\alpha_i=\Delta\omega^2/(2a)$ we obtain together with $\alpha_r/\alpha_i =1/a$ the final approximate expression
\be
{\cal I}_m \approx
\frac{2\pi}{\Delta\omega^2}
\Theta(-\eta_m a)
\exp\left(-\eta_m^2\right)
\exp\left(i\,\eta_m^2 a\right).
\ee
We substitute this formula  into \eq{eUg3} for the probability amplitude $c_e$
and arrive at \eq{ce} with the approximate weight factor
\be
w_m=-d_m \Theta(-\eta_m a)
\exp\left(-\eta_m^2\right).
\label{weight:theta}
\ee
This expression also results from the exact formula \eq{eq:weight:1} with the help of the 
relation $\sqrt{1-i a}\stackrel{\sim}{=} (-i)^{1/2}a$ together with the
primitive asymptotic expansion 
\be
\text{erfc}(\zeta_m)\approx 2\Theta(-\eta_m\, a)
\ee
of the complementary error function following from Eqs.~(\ref{erg1}) and (\ref{erg2}).

\subsubsection{Interference in time-frequency phase space}

\begin{figure}
\begin{center}
\includegraphics[scale=0.5]{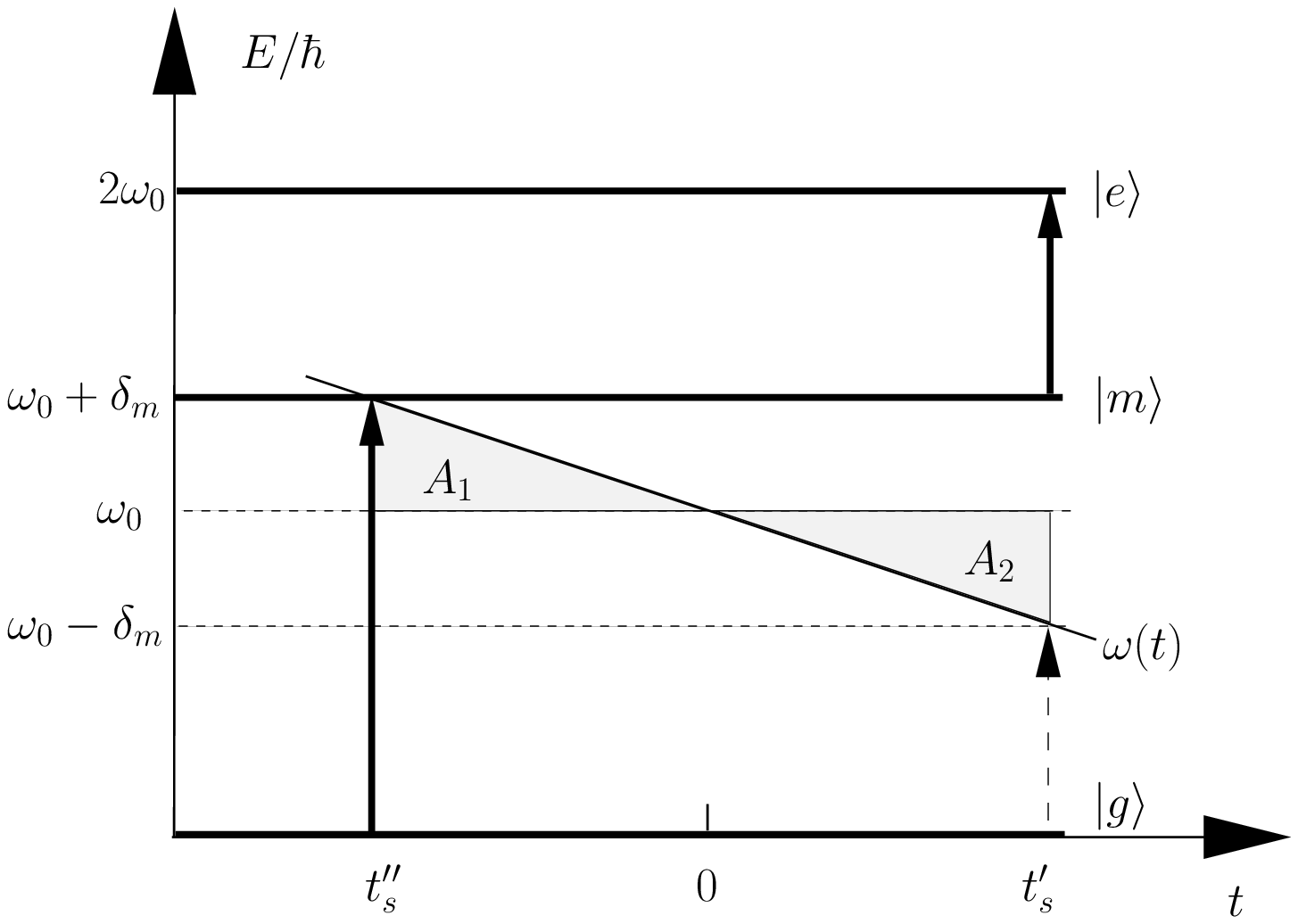}
\end{center}
\caption{Origin of Gauss sum from phase space interpretation of two-photon excitation induced by a chirped pulse.
In the phase space spanned by frequency $\nu$ and time $t$ we indicate the ground state $\ket{g}$, the intermediate level $\ket{m}$ and the excited state $\ket{e}$ of frequencies $0,\,\omega_0+\delta_m$ and $2\omega_0$, respectively, by thick horizontal lines. 
We also denote the one-photon resonance of energy $\omega_0$ and the virtual level of frequency $\omega_0-\delta_m$  by dotted horizontal lines.
The instantaneous frequency $\omega(t)=\omega_0+2\alpha_i\,t$ is depicted for one negative value of the chirp by the tilted thin line. A stationary phase analysis of the double integral \eq{I:m} determining the two-photon transition probability amplitude brings out most clearly that the sequential excitation appears at the times $t_s''\equiv \delta_m/(2\alpha_i)$ and $t_s'$. Indeed, at $t_s''$ the instantaneous frequency is resonant with the transition from $\ket{g}\to\ket{m}$ whereas at $t_s'$ it is resonant with the transition from $\ket{m}$ to $\ket{e}$ as indicated by the dotted and the thick arrow. In this phase space representation the times when the transitions occur are determined by the crossings of the tilted thin line with the thick and the lower dotted  horizontal lines corresponding to the frequencies $\omega_0+\delta_m$ and $\omega_0-\delta_m$. The enclosed shaded area $A_1+A_2$ corresponds to the phase acquired during the transition. Each triangle  has the area $\delta_m t_s''/2= \delta_m^2/(4\alpha_i)$. As a consequence, the phase associated with this transition is quadratic in the offset $\delta_m$. In the case of a multi-state ladder with a harmonic manifold we have the interference of all excitation paths, that is the weighted sum of phase factors with quadratic phases giving rise to a Gauss sum.}
\label{3state}
\end{figure}

This analysis of the transition probability amplitude in second order perturbation theory translates itself into an elementary geometrical representation in a phase space spanned by frequency $\nu$ and time $t$.  In \fig{3state} we consider a single sequential excitation path in this space. We indicate the frequencies  $0,\,\omega_0+\delta_m$ and $2\omega_0$ of the ground state, the $m$-th intermediate state and the excited state by thick horizontal lines.  We also denote the frequency $\omega_0$ of the two-photon resonance and the virtual level of frequency $\omega_0-\delta_m$  by dotted horizontal lines. The instantaneous frequency $\omega(t)=\omega_0+\alpha_i\,t$ is depicted for one negative value of the chirp by the tilted thin line. 

The transitions from $\ket{g}$ to $\ket{m}$ and from $\ket{m}$ to $\ket{e}$ occur at times $t_s''$ and $t_s'$, respectively. These times are determined geometrically by the crossings of the representations of the instantaneous frequency $\omega(t)$ and the two frequencies $\omega_0+\delta_m$ and $\omega_0-\delta_m$, that is by the crossings of the tilted thin line with the thick and the lower dotted horizontal lines. Each crossing contributes to the total transition probability amplitude with a phase determined by the enclosed triangular shaded area $A_1+A_2$ corresponds to the phase acquired during the transition. Each triangle  has the area $t_s'' \delta_m/2= \delta_m^2/(2\alpha_i)$. As a consequence, the phase associated with this transition is quadratic in the offset $\delta_m$.

\subsection{Direct excitation}
We conclude by briefly discussing the direct two-photon excitation.
In the analysis of the time ordered integrals \eq{int:timeorder} we have implicitly assumed that the moments $t_s'$ and $t_s''$ of the transitions are appropriately separated in time. However, since we deal with a double integral the instant $t'=t''$ can make a significant contribution. This fact stands out most clearly in the re-formulation, \eq{timechange}, of the double integral where the line $t'=t''$ translates into the condition $t=0$. When we recall the definition \eq{subst} of the path ${\cal C}(\zeta)$ of integration the moment $t=0$ translates into the starting point $\zeta_m$ of the path. As a consequence the main contribution to the integral $I_2$ arises from the lower limit. The partial integration that we have used in Sec.~\ref{asymp} to derive the asymptotic expansion of the complementary error function took advantage of this fact.

\section{Manifold of intermediate states}
\label{manifold}
So far we have concentrated on a single intermediate state. We now briefly address the problem of a manifold of intermediate states.
In the case of a multi-state ladder with a harmonic manifold we have the interference of all excitation paths, that is the weighted sum of phase factors with quadratic phases giving rise to a Gauss sum. In the case of a positive chirp parameter $a$ these paths consist solely of direct two-photon excitations. In the limit of negative values of $a$ each interfering path through the $m$-th level consists of an interference of a sequential and a direct excitation. However, the direct excitation is less pronounced especially for appropriately negative values of the chirp. In this case the probability amplitude is predominantly determined by the interference of the quadratic phases arising from the sequential excitations.

\subsection{Gauss sum made explicit}

In order to bring out most clearly the connection to the Gauss sums, we make 
the dependence of the weight factors and the phase  on $m$ more explicit. For this purpose we use the definitions, \eq{energy} and \eq{scaledchirp} for the offset $\delta_m$ of the harmonic manifold and the second order dispersion $a$ to verify the quadratic nature of the phase  
\be
\eta_m^2\,a=\left(\frac{\delta_m}{\Delta\omega}\right)^2\,a
=\frac{\pi}{2}N \xi+2\pi\left(m +\frac{m^2}{N}\right)\xi
\label{phase:factor}
\ee
in the probability amplitude $c_e$. Here we have introduced the quantity
\begin{equation}
N\equiv\frac{2\delta}{\Delta}
\label{number}
\end{equation}
and the dimensionless chirp
\begin{equation}
\xi\equiv\frac{\delta\Delta}{\pi} \frac{a}{\Delta\omega^2} =\frac{\delta\Delta}{\pi} \phi''.
\label{argument}
\end{equation} 
Moreover, in terms of  $N$  the offset $\delta_m$ defined by \eq{energy} reads 
\be
\delta_m
=\Delta\left(m+\frac{\delta}{\Delta}\right)
=\Delta\left(m+\frac{N}{2}\right).
\label{delta:N}
\ee
Hence, the final expression for the excited state probability amplitude takes the canonical form
\begin{equation}
c_e=\exp\left(i\, \frac{\pi}{2}N \xi\right)\,
\sum\limits_{m=-M'}^M w_m \,
\exp\left[2\pi i\left(m+\frac{m^2}{N}\right)\xi\right]
\label{truncated:gauss}
\end{equation}
of a Gauss sum. Here we have numbered the intermediate levels from $-M'$ to $M$.

When we express $\delta_m$ with the help of \eq{delta:N} the weight factors
\be
w_m\equiv-\frac{d_m}{2}\, 
\text{erfc}\left(\zeta_m \right)
\exp\left[-\left(\frac{m+N/2}{\Delta m}\right)^2\right]
\label{eq:weight}
\ee
as well as the argument 
\be
\zeta_m\equiv
\frac{m+N/2}{\Delta m}i\sqrt{1-ia}
\ee
of the complementary error function contain the width $\Delta m \equiv \Delta\omega/\Delta$. 

The transition probability amplitude consists of a sum of weighted phase factors with phases that depend linearly and quadratically on the summation index $m$. Closely related sums appear in the context of fractional revivals\cite{leichtle:PRL:1996,leichtle:PRA:1996}, the Talbot effect\cite{schleich:2001} and curlicues\cite{berry:curlicues:1:1988,berry:curlicues:2:1988}
and Josephson junctions\cite{schopohl}.

\subsection{Asymptotic expansion of weight factors}
\label{approx:weight}
With the help of the asymptotic expansion \eq{weight:theta} the approximate real-valued weight factors
\be
w_m \approx -d_m\, \Theta\left[-a\,\left(m+\frac{N}{2}\right)\right] \exp\left[-\left(\frac{m+N/2}{\Delta m}\right)^2\right]
\label{weights:approximated}
\ee
are given by half a Gaussian centered around $m \approx -N/2$. The sign of the chirp determines which half. Indeed, for negative chirps that is $a<0$ we find only contributions for $m>-N/2$. In contrast for positive chirps, that is for $0<a$ only levels with $m<-N/2$ contribute.
Moreover, by controlling the width $\Delta m$ we can ensure that all quantum paths enter the sum with approximately the same weight. 

\begin{figure}[ht]
\begin{center}
\includegraphics[width=0.9\columnwidth]{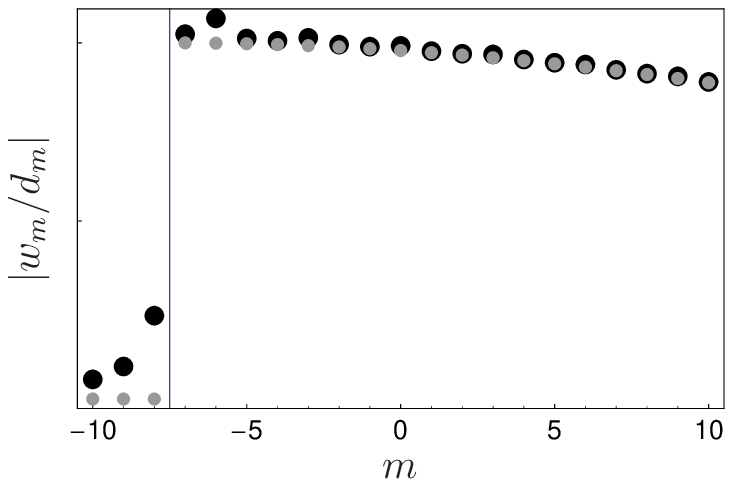}
\end{center}
\caption{
Comparison between the exact (black dots) and the approximate (gray dots) expression for the scaled weight factors $|w_m/d_m|$ given by Eqs.~(\ref{eq:weight}) and (\ref{weights:approximated}), respectively. 
Here we consider the case  $N=15$ with $\Delta m=\Delta\omega/\Delta=50.8$
for the negative dimensionless chirp $a=-10824<0$.
}
\label{weight}
\end{figure}

In \fig{weight} we compare and contrast the exact (black dots) and the approximate (gray dots) expressions, Eqs.~(\ref{eq:weight}) and (\ref{weights:approximated}), for the  ratio  $|w_m/d_m|$ of  the weight factors $w_m$ and the matrix element $d_m$. To be specific, we have used $N=15$ and a negative chirp. The figure clearly shows that in this case the weight factors are non-vanishing only on the right half of the Gaussian, that is for $-N/2=-7.5<m$. Moreover, there is an excellent agreement between the exact result and the approximation.

We conclude by noting that there are small oscillations in the exact weight factor distribution which are not contained in the approximate expression. They are a remnant of the interference between the direct and the sequential path as discussed in Sec.~\ref{interference}.

\section{Conclusion and outlook}
\label{conclusion}
In the present paper we have analyzed a two-photon transition in a ladder system driven by a chirped laser pulse. We have shown that the resulting excitation probability amplitude is of the form of a Gauss sum.
However, one task remains: Find a quantum system which displays such an equidistant ladder system. 

Before we address this task it is helpful to recall the origin of the Gauss sum -- the interference of paths with quadratic phases. \fig{3state} brings out most clearly that these phases arise already from the first step of the excitation, that is from  transition into the equidistant manifold. Therefore,  a one-photon transition from a ground state to an equidistant manifold of excited states driven by a chirped pulse also leads to a Gauss sum.
In this simplified system the fluorescence signal is acquired from the states in the harmonic manifold. Interference from the individual levels occurs during spontaneous emission of the states. 

At least five different quantum systems offer themselves as candidates for such one-photon excitations: 
({\it i}) Rydberg atoms in electric or magnetic fields, 
({\it ii}) vibrational energy levels in a molecule, 
({\it iii}) atoms in traps, 
({\it iv}) laser driven one-photon transitions, and
({\it v}) quantum dots.
However, these systems have to be subjected to additional requirements: ({\it i}) the Rabi-frequencies for the interfering excitation paths should ideally be of the same order of magnitude in order to guarantee that they contribute in a democratic way, that is with the same weight, ({\it ii}) since anharmonicities in the target manifold change the character of the resulting probability amplitude, only perfectly equidistant states result in  a Gauss sum.
We now briefly discuss these quantum systems and address these requirements

Rydberg atoms\cite{gallagher:1988} are highly sensitive to electric or magnetic fields. In an electric field the energy levels of a Rydberg atom split into a manifold of states\cite{Kleppner:1979}.
The so-called Stark maps which show the dependence of the splitted spectral lines as a function of the electric field have been measured for  many alkali atoms. Whereas for a modest field strength the separation between neighboring splitted energy levels is constant, for larger values of the field also higher order corrections to the energy have to be taken into account.
A closer analysis of the Stark map reveals that the dipole moments vary for each member of the manifold and the individual paths do not contribute equally. Due to the Zeeman effect a homogeneous magnetic field also creates  a  manifold of equidistant states $\ket{m}$. However, in this case the problem are the selection rules which eliminate most of the transitions. A change of the orientation of the magnetic field  might offer a possibility to overcome this problem.

Diatomic molecules may also serve as a candidate system for identifying a harmonic manifold of (almost) equidistant states\cite{wallentowitz:2002}. 
The diagonalization of the ro-vibrational Hamiltonian yields 
that the inter-nuclear potential in a  diatomic molecule can be indeed approximated as a harmonic oscillator with equidistant spectrum. However, anharmonicities have again to be taken into account and would destroy the regularity of the Gauss sum.

Another promising realization of an equidistant spectrum relies on cold trapped atoms or ions\cite{phillips:rmp:1998,wineland:rmp:1999}. The central idea of this approach is that the trapping in  magnetic microtraps relies  on the magnetic dipole of the atom which depends on its internal state.
Therefore, the center-of-mass motion of an atom depends on its internal state.
Our suggestion for the realization of an equidistant spectrum is reminiscent of the quantum tweezer\cite{raizen:PRL,mohring}.
Whereas the atom initially in the ground state is weakly bound to a shallow trap, the atom in its excited state feels a steep harmonic potential. The vibrational states in the excited electronic state constitute the desired harmonic manifold. 

Yet another method to obtain an equidistant ladder system is to
apply two electromagnetic fields to a two-level atom which has a permanent dipole moment in the excited state. A strong cw-modulation field generates an equidistant Floquet ladder in the excited state. The second field, that is the weak chirped pulse, leads to the excitation probability amplitude in the form of a Gauss sum. 

Quantum dots\cite{ashoori:1996} may be the perfect choice in the end. Indeed, they allow for designing the discrete energy spectrum\cite{kouwenhoven:1998} of a trapped electron and can thus be viewed as an artificial atom. Quantum dots have already been discussed as a candidate system for implementing quantum logic\cite{divincenzo:1998}.

In conclusion, we have shown that a two-photon transition in a ladder system driven by a chirped laser pulse leads to an excitation probability amplitude in the form of a Gauss sum. We have discussed several physical systems in the context of realizing such an excitation scheme and have also addressed the experimental challenges. We emphasize that this example also establishes an interesting connection between number theory and wavepacket dynamics. 

\begin{acknowledgements}

We thank I.~Sh.~Averbukh, M.~Bienert, P.~Braun, P.~T.~Greenland, F.~Haug, P.~Knight, T.~Pfau, F.~Schmidt-Kaler, S.~Wallentowitz and A.~Wolf for stimulating discussions.
W.~M., E.~L. and W.~P.~S. acknowledge financial support by 
the Landesstiftung Baden-W\"{u}rttemberg. In addition, this work was  supported by a grant from the Ministry of Science, Research and the Arts of Baden-W\"urttemberg in the framework of the Center of Quantum Engineering. Moreover, W.~P.~S. also would like to thank the Alexander von Humboldt Stiftung and the Max-Planck-Gesellschaft for receiving the Max-Planck-Forschungspreis.
\end{acknowledgements}


\begin{thebibliography}{36}
\bibitem[{chi({\natexlab{b}})}]{chirp:footnote:2}
\bibinfo{note}{We note that this system is reminiscent of the phenomenon of
double optical resonance\cite{greenland} emerging in a pump-probe experiment in a three-state ladder. The pump transition is driven by a strong pulse such that the intermediate state experiences a time-dependent shift due to the Autler-Townes effect. The probability amplitude of exciting the upper state is the superposition of two contributions arising from the instances of time when the weak probe pulse is on resonance with the ac-Stark shifted intermediate level.}
\end{thebibliography}

\end{document}